\documentclass[12pt]{article}
\usepackage{graphicx}

\begin{document}

\title{On the surface critical behaviour in Ising strips: 
density-matrix renormalization-group study}
\author{A. Drzewi\'nski \\
Czestochowa University of Technology, \\
Institute of Mathematics and Computer Science,\\
ul.Dabrowskiego 73, 42-200 Czestochowa, Poland\\*[4mm]
A. Macio\l ek \\
Institute of Physical Chemistry, Polish Academy of Sciences,\\
Department III, Kasprzaka 44/52, PL-01-224 Warsaw, Poland\\*[4mm]
K. Szota \\
Czestochowa University of Technology, \\
Institute of Mathematics and Computer Science,\\
ul.Dabrowskiego 73, 42-200 Czestochowa, Poland}

\maketitle

\begin{abstract}
Using the density-matrix renormalization-group method we study the surface critical behaviour 
of the magnetization in Ising strips in the subcritical region. Our results support the prediction
that the surface magnetization in the two phases along the pseudo-coexistence curve also
behaves as for the ordinary transition below the wetting temperature for the finite value of
the surface field.
\end{abstract}

PACS numbers: 05.50.+q, 68.35.Rh, 68.05.Bc

\section{Introduction}
\label{sec:int}
More than two decades of recent studies have yielded a fairly detailed
understanding of the critical behavior at surfaces \cite{binder:83:0,diehl:86:0}.
However, attempts to verify  theoretical predictions,
both in experiments and in model systems, often  point out
issues which need further clarifications.
We came across such an issue in the recent work by Brovchenko et al.,
where the surface critical behavior of  a model water in the slitlike
and cylindrical  pores~\cite{brovchenko:04}, and of
a Lennard-Jones fluid in  the slitlike pores~\cite{brovchenko:05}
was  studied by means of Monte Carlo simulations. In both cases
fluid particles were assumed to interact with a wall via a (10-4)
long-range potential and a parameter that
measures the well depth of the wall-fluid potential was chosen 
to correspond to a weakly attractive surface.
A one-component fluid like water or the Lennard-Jones fluid  is
expected  to lie in the universality
class of the {\it  normal}  surface transition of semi-infinite
Ising system. In a magnetic  language  the normal transition
is characterized by  {\it two} relevant scaling fields,
a surface scaling field $c>0$  and  a non-zero  external
surface field $|h_1|>0$.
$c$  describes the enhancement of interactions in the surface layer.
$c>0$ corresponds to a reduced tendency to order in the surface,
which is the case generic for fluid systems because  the
presence of a wall should decrease the net fluid-fluid attraction
between a molecule and its nearest neighbors below the
corresponding bulk value. On the other hand,
the containing walls exert an effective potential on a fluid and in
magnetic language this  coresponds to some generally
{\it nonzero} surface field $h_1$.
There is a possibility to mimic the situation  of vanishing
surface field $h_1=0$, i.e., the so called  {\it ordinary} surface transition
behavior,
by  suitable tunning wall-fluid interactions relative to
fluid-fluid interactions. Due to the lack of Ising symmetry in
a ``real'' fluid, it is very unlikely to find a wall-fluid
potential that  corresponds exactly to $h_1=0$, however one does  find
a wall-fluid potential which is ``neutral''. As was shown in 
Ref.~\cite{maciolek:03} for $T\ge T_c$, this ``neutral'' wall gives 
rise to the Gibbs
adsorption $\Gamma \sim 0$ that is constant along the critical isochore
and is characterized by a fluid density profile which,
away from the walls where oscillations arise,
is almost flat throughout the slit~\cite{maciolek:03}.
For the ``neutral'' wall a parameter that
measures the well depth of the wall-fluid potential corresponds
to a weakly attractive surface. 
In Refs.~\cite{brovchenko:04,brovchenko:05} even more weakly attractive 
substrates were considered. The authors focused
on the subcritical regime and studied the temperature dependence of 
density profiles along the pore liquid-vapour coexistence curve.
Recall, that the normal transition is governed by the fixed point of the
renormalization-group transformation $h_1=\infty, c=\infty$  and should be
equivalent to the extraordinary transition given by fixed point
$h_1=0, c=-\infty$~\cite{bray:77,burkhardt:94,diehl:94}.
At these (equivalent) surface transitions, the order parameter (OP)
at the surface layer $m_1$ should have a leading thermal
singularity of the same form as the bulk free energy,
i.e., $|T_c-T|^{2-\alpha}$, where $T_c$ is bulk critical temperature.
More precisely one expects for $\tau\to 0$ ~\cite{bray:77,burkhardt:94,diehl:94}
the limiting behavior
\begin{equation}
\label{eq:ex}
m_1-(m_{1C}+A_1\tau+A_2\tau^2+\ldots)\approx +A_{2-\alpha}^{\pm}|\tau|^{2-\alpha},
\end{equation}
where $\tau\equiv (T_c-T)/T_c$, and the contribution in parentheses
is a regular background.
For both  model fluids Brovchenko et al.~\cite{brovchenko:04,brovchenko:05}
defined the local OP  as
$\Delta \rho(z)\equiv (\rho_l(z)-\rho_v(z))/2$, where $\rho_l(z)$ and
$\rho_v(z)$ are the  density profiles of the coexisting liquid and
vapour phases, respectively, and  found that below
the bulk critical temperature $T_c$ this OP  shows the behavior
which is in accordance with the  ordinary transition.
In particular, near the surface  a variation of $\Delta \rho$
with reduced temperature $\tau$
follows the scaling law with a value of the exponent close to the
$\beta _1\simeq 0.82$
of the ordinary transition in the Ising system in d=3, i.e.
\begin{equation}
\label{eq:ord}
\Delta \rho_1(\tau) \approx \tau^{\beta _1}.
\end{equation}
On the basis of these  observations, made for the confined fluids,
the authors put forward
the hypothesis that the difference $\Delta \rho$ between
the densities of coexisting phases near the surface should follow the
behavior (\ref{eq:ord})  also near strongly
attractive  surfaces below the wetting temperature $T_w$.
This is based on the assumption
that the term $\sim \tau^{\beta_1}$ should always be present in both
coexisting phases below $T_w$.
The authors reconsider the
surface critical behavior of the semi-infinite Ising model claiming that
below the wetting  temperature  the surface magnetizations $m_1^{I}$
and $m_1^{II}$ in the two phases along the coexistence curve should have the
following limiting behavior for $\tau\to 0$:
\begin{eqnarray}
\label{eq:mag}
m_1^{I}&=&B_1\tau^{\beta_1}+m_{1C}+A_1'\tau+\ldots +A_{2-\alpha}^-|\tau|^{2-\alpha} \\ 
m_1^{II}&=&-B_1\tau^{\beta_1}+m_{1C}+A_1'\tau+\ldots +A_{2-\alpha}^-|\tau|^{2-\alpha}
\end{eqnarray}
The symmetric term $\sim \tau^{\beta_1}$,
which describes the temperature dependence of the magnetization at $h_1=0$, 
accounts for the  missing-neighbor effect and, as the authors claim, was 
overlooked in Ref.~\cite{burkhardt:94,diehl:94}.
Above the wetting temperature there exist  a single phase
 which is expected to have the surface magnetization of  the form given by the above equation.

For $d=2$ semi-infinite Ising model there exist  exact results for $m_1$ in
the presence of the surface field. They were derived by McCoy and Wu and also
by Au-Yang and Fisher using a Pffafian method \cite{mccoy:67,au-yang:80}. Specifically,
let us consider a planar rectangular lattice with coordinates $i$ (horizontal)
and $j$ (vertical) with spins $\sigma_{i,j}=\pm 1$ located at the
sites of the lattice and  interacting with nearest-neighbors via the
coupling $K=\beta J>0, \beta=1/k_BT$. Assume
vanishing bulk magnetic field $h=0$, a cycling boundary
condition in the horizontal direction
and a surface magnetic field $h_1$, measured in the units of the
coupling constant $J$, interacting  with one of the two
horizontal boundary rows of spins. On the second boundary the spins are free.
The configurational Hamiltonian  is defined as
\begin{equation}
\label{eq:int0}
{\cal H} =-J\sum_{i,j}\sigma  _{i,j}\sigma _{i,j+1}-
J\sum_{i,j}\sigma  _{i,j}\sigma _{i+1,j}-h_1\sum_{j}\sigma _{1,j},
\end{equation}
where the sums run over $1\le i\le M$ and $1\le j \le N$.
The analytic expression for the  free energy of this system
was obtain as a sum \cite{mccoy:67}
\begin{equation}
\label{eq:mac1}
MNF+2N{\cal F}_0+N{\cal F}.
\end{equation}
The first term is the bulk free energy, and the
terms $2N{\cal F}_0$ and $N{\cal F}$
are additional contributions coming from the existence of the free boundary
which interacts with the surface magnetic field.
The magnetization $m_1$ of the first row
was calculated from
\begin{equation}
\label{eq:mac2}
m_1=-\frac{\partial {\cal F}}{\partial h_1}
\end{equation}
and various limiting cases were discussed.
In the case relevant for the ordinary transition,
i.e., for $h_1\to 0$, the boundary spontaneous magnetization
exists only in the thermodynamic limit  $M,N\to \infty$:
\begin{equation}
\label{eq:mac3}
m_1(0^+)=\lim _{h_1\to 0} m_{1}(h_1)=\left[ \frac{\cosh 2\beta J-\coth 2\beta J}{\cosh 2\beta J-1}\right]^{1/2}.
\end{equation}
This  vanishes  at the critical temperature as
\begin{equation}
\label{eq:mac4}
 m_1(0^+)\approx \left[\frac{2\ln (1+{\sqrt 2})}{\sqrt{ 2}-1}\right]^{1/2}|\tau|^{1/2}
\end{equation}
from which one can read off the value of the surface critical exponent
$\beta^{ord}_1$ for
$d=2$ Ising model, i.e.,  $\beta^{ord}_1=1/2$.
The case when $T$ is near $T_c$ and $h_1$
is positive and {\it away} from zero is relevant for the normal transition.
For this case the  result is:
\begin{equation}
\label{eq:mac6}
m_1(h_1)=\textrm{ Taylor ser. in}\quad \tau + \\
\frac{(\sqrt{2}-1)(1-z_1^2)}{\pi z_1^2}\left(\frac{2J}{\beta_c}\right)^2\tau^2\ln|\tau|,
\end{equation}
where $ z_1=\tanh \beta h_1$.
Thus the leading singularity of the boundary magnetization
agrees with the prediction by Diehl \cite{diehl:94} for the normal
surface transition, i.e., $m_1$ has a leading thermal singularity
of the same form $\tau^{2-\alpha}$ as the bulk free energy. $\alpha=0$ in
$d=2$ Ising model which corresponds to the logarithmic behavior.
In Ref.~\cite{au-yang:80} the first two terms of the Taylor series in $\tau$
were given explicitly
\begin{equation}
\label{eq:mac6a}
m_1(h_1)=m_{1,c}(h_1)+D_1(z_1)\frac{2K_c\tau}{z_1}+O(\tau^2/z_1^3),
\end{equation}
where
\begin{eqnarray}
\label{eq:mac6b}
\frac{1}{2}\pi D_1(z_1) &=& 1+(1+{\sqrt 2})^2z_1^2\ln z_1^2 \nonumber \\
+& (1+{\sqrt 2})^2&\left[\frac{1}{4}\pi+{\sqrt 2} + \ln (1+1/{\sqrt 2})\right]z_1^2.
\end{eqnarray}
On the other hand, if $h_1$ is nonzero but {\it  small} then,
for $T$ near $T_c$, the limiting behavior of $m_1$ is different:
\begin{equation}
\label{eq:mac7}
m_1\sim \frac{(1-\alpha)^{1/2}}{z^{1/2}}-\frac{2z_1}{\pi z}\ln(1-\alpha+z_1^2),
\end{equation}
where $\alpha=(1-z)/z(1+z)$, $z=\tanh K$ and
$\alpha = 1-4K_c\tau +O(\tau^2)$ for $\tau\to 0$. Thus, as $z_1\to 0$
Eq.~(\ref{eq:mac7}) agrees with Eq.~(\ref{eq:mac3}) and
exhibits the square-root behavior of the ordinary transition.
These exact results show explicitely that 
for strong surface field  the prediction (\ref{eq:mag})
is not  true  sufficiently close to the 
critical temperature. On the other hand  in view of Eq.~(\ref{eq:mac7})
it is  understandable why simulation results for weakly attracting
substrates \cite{brovchenko:04,brovchenko:05}
may show the ordinary transition behavior. However, results described 
above concern the behavior of the boundary magnetization
only for one of the two possible bulk phases, and the temperature dependence
of the {\it difference} between the magnetizations of both coexisting phases 
near the surface has not been studied. This is due to the fact
that in the absence of the bulk magnetic field the choice of the sign of $h_1$
breaks the symmetry in the finite system, and, for example, the positive surface field
yields $(+)$ phase in the bulk in the thermodynamic limit.
In order to calculate the boundary magnetization for the case of the $(-)$
bulk phase in the presence of the positive boundary field, one would have to
perform calculations in the presence of infinitesimaly
small negative bulk field and put $h\to 0^-$ after taking the thermodynamic
limit or to solve the model with very sophisticated boundary conditions.

\begin{figure}[h]
\centering
\includegraphics[width=8.0cm]{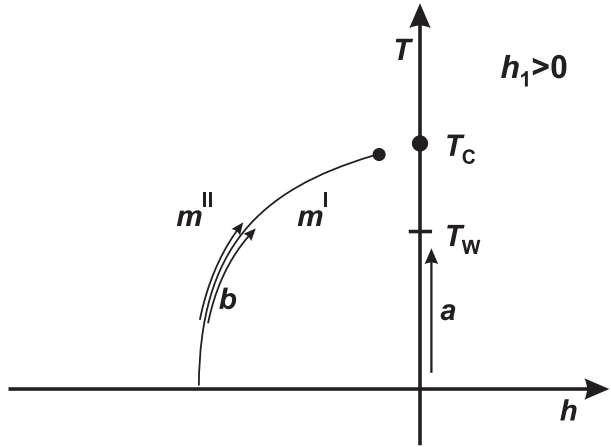}
\caption{Schematic phase diagram for an Ising strip with positive surface
fields. There are two thermodynamic paths presented: a) along the bulk
coexistence and b) along the (pseudo)coexistence of the confined system.}
\label{fig1}
\end{figure}

So far exact solution of the $d=2$ Ising model at nonvanishing
bulk magnetic field is not available, however, recently
developed  density-matrix renormalization-group (DMRG)
method~\cite{dmrg} allows for very accurate numerical calculations
in the presence of the {\it arbitrary} surface and bulk fields.
The DMRG method, based on the transfer matrix approach for calculating the
partition functions, includes the critical fluctuations and therefore is
very suitable for studying the critical behavior. In the following we will
use this approach to calculate the magnetization profiles in the strip
geometry, i.e., the geometry analogous to the one studied in 
Refs.~\cite{brovchenko:04,brovchenko:05}. We assume identical surface fields
$h_1=h_2>0$ acting on the two boundaries separated by a finite distance
$L$ and consider two different
thermodynamic paths: (i) along the bulk coexistence  $h=0$, and (ii)
along the pseudo-coexistence of the confined system $h=h_{co}(T)$
(see Fig.\ref{fig1}).
The first path is the one for which exact results summarized
above have been obtained.  In this case we want to explore how 
the finite-size effects in the confined geometry influence
the crossover from the one type of the surface critical behavior to another.
The second path allows  to study  the temperature dependence
of the {\it difference} between the surface magnetizations of both
pseudo-coexisting phases.

The DMRG provides an efficient algorithm to construct the effective
transfer matrix $\cal T_L$ for large two-dimensional classical systems
at finite temperatures \cite{DMRG}.
Starting with a small system (e.g. $L=4$ in our case), for which $\cal T_L$
can be
diagonalized exactly, one adds iteratively couples of spin columns until
the allowed
(in the computational sense) size of the effective matrices is reached.
Then further
addition of new spins forces one to discard simultaneously the least important
states to keep the size of the effective transfer matrices fixed.
This truncation is done through the construction of a reduced density
matrix whose eigenstates provide the optimal basis set $m_{\lambda} $.
The size of the effective transfer matrix is
then substantially smaller than the original dimensionality of the
configurational space $(2m_{\lambda})^2 \ll 2^L$. Generally,
the larger is $m_{\lambda }$, the better
accuracy is guaranteed. In the present case, we keep this parameter up to
$m_{\lambda}=40$. Typically a truncation error was not larger than $10^{-12}$.
We estimate that the errors in the plots are smaller than the symbol size.
The DMRG method allows to study much larger systems (up to $L=340$ in this 
paper) than it is possible with
standard exact diagonalization method which can handle with systems up to
several dozens columns for Ising model. Comparisons with exact results
for the case of vanishing bulk magnetic field show that this technique gives
very accurate results in a wide range of temperatures \cite{Maciolek}.

\begin{figure}[h]
\centering
\includegraphics[width=8.0cm]{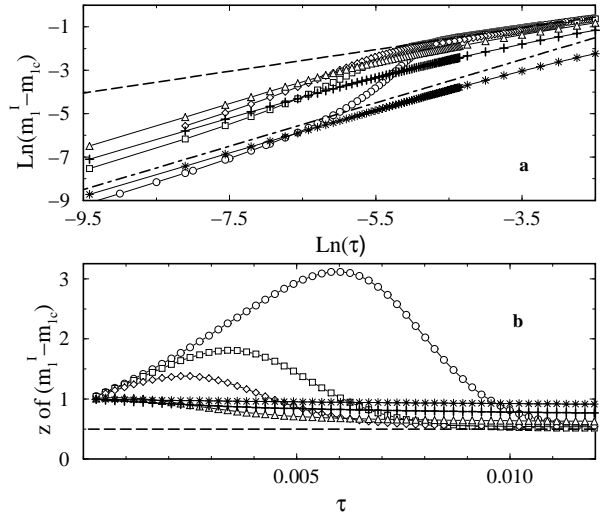}
\caption{The difference ($m_{1}^{I}$ - $m_{1c}^{I}$) as a function of $\tau$ at $L=340$ for
various surface fields (circles $h_1$=0.001, squares $h_1$=0.005, diamonds $h_1$=0.01,
triangles $h_1$=0.03, crosses $h_1$=0.1, stars $h_1$=0.5): a) the log-log plot
b) the effective exponent. The dashed line denotes the slope 1/2 and the dotted-dashed
line describes the slope 1.
}
\label{fig2}
\end{figure}

\begin{figure}[h]
\centering
\includegraphics[width=8.0cm]{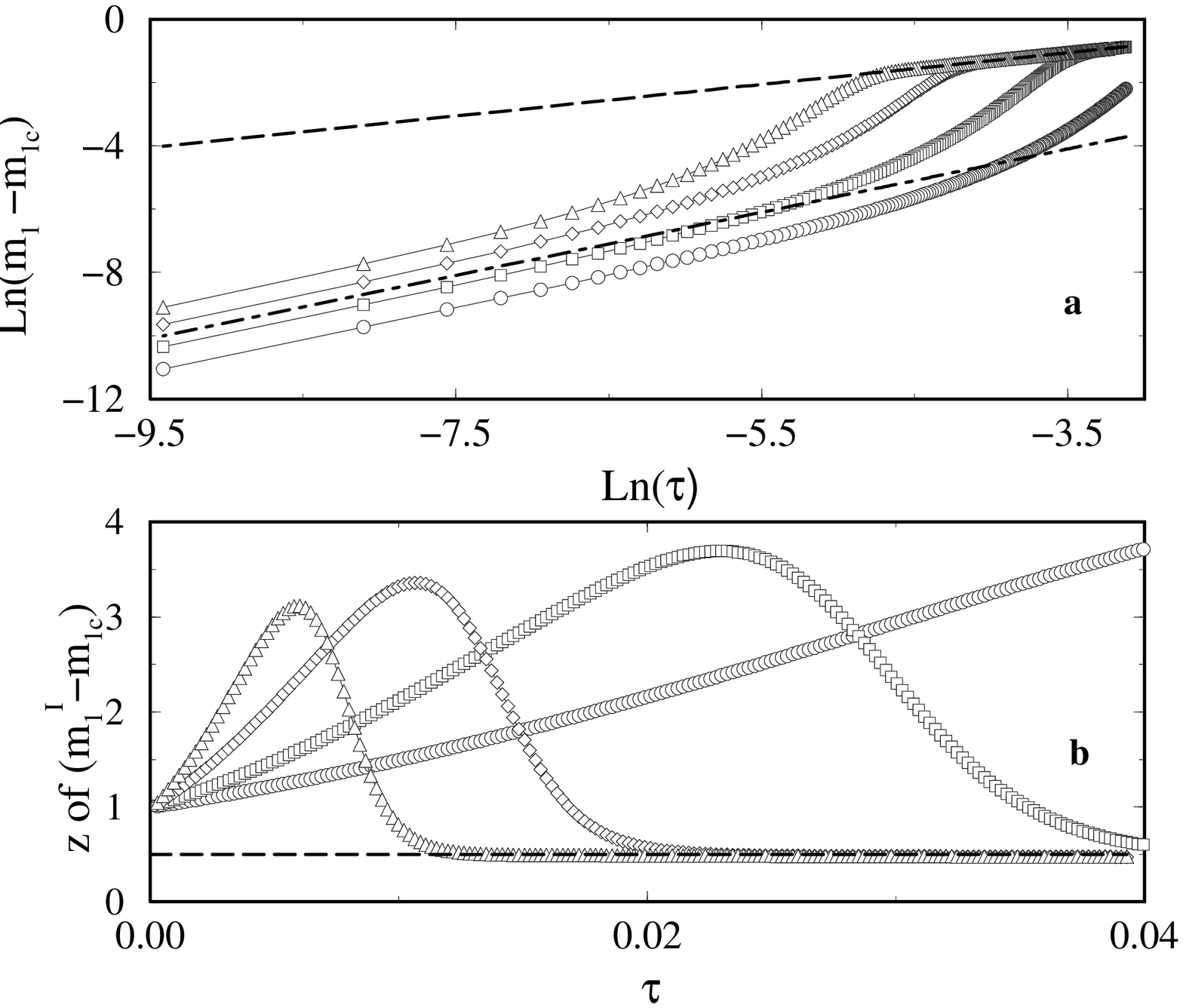}
\caption{The difference ($m_{1}^{I}$ - $m_{1c}^{I}$) as a function of $ \tau$ 
at $h_1$= 0.001 for various strip widths $L$ (circles $L=$50, squares $L=$100, 
diamonds $L=200$, triangles $L=340$): a) the log-log plot b) the effective exponent.  
The dotted-dashed line denotes the slope 1, whereas the dashed describes 
the slope 1/2.
}
\label{fig3}
\end{figure}

\section{Along the bulk coexistence $h=0$}
\label{sec:1}

First we discuss results for $h=0$.
Recall, that in a finite system with the positive $h_1$ and $h=0$,
below $T_c$ there exist only a $(+)$ phase characterized
by  magnetization profiles $m^I(z)$  which are  positive across the
strip~\cite{fisher:1}.
In the Fig.~\ref{fig2}a we show the log-log plot of the difference
$m^I_1-m^I_{1c}$ as a function of $\tau$  calculated for the strip
of the width $L=340$ and for the  selection of the surface fields.
For the weakest considered surface fields,
i.e., for  $h_1 =0.001, 0.005, 0.01$, and 0.03,
we find the square-root behavior of the
ordinary transition but only for  temperatures  $\tau > 0.01$.
In an  agreement with the exact result (\ref{eq:mac7}) the amplitude of
this leading decay  does not depend on $h_1$.
For smaller $\tau$ there is a  crossover to the linear behavior with the
$h_1$-dependent  amplitude. The range of temperatures in which the crossover
takes place depends sensitively on the value of the surface field, the weaker
$h_1$ the further away from $T_c$ the crossover starts, but in any case the linear 
behavior is observed for $\tau < 0.001$. This is very well illustrated in
the plot of the {\it effective exponent} of  $m^I_1-m^I_{1c}$
as a function of $\tau$ (Fig.~\ref{fig2}b).
The effective exponent, i.e. the following quantity
\begin{equation}
z_i = \frac{\ln m_1 (i+1) - \ln m_1 (i)}{\ln \tau (i+1) - \ln \tau (i)} \,\, ,
\label{zN}
\end{equation}
is the discrete derivative of the data in the log-log scale plot.
Such quantity probes the local slope
(at a given reduced temperature $\tau (i)$)
providing a better estimate of the leading exponent than a log-log plot.
The calculation of $z_i$ requires very accurate data that can be quaranteed
by DMRG data. 

\begin{figure}[h]
\centering
\includegraphics[width=8.0cm]{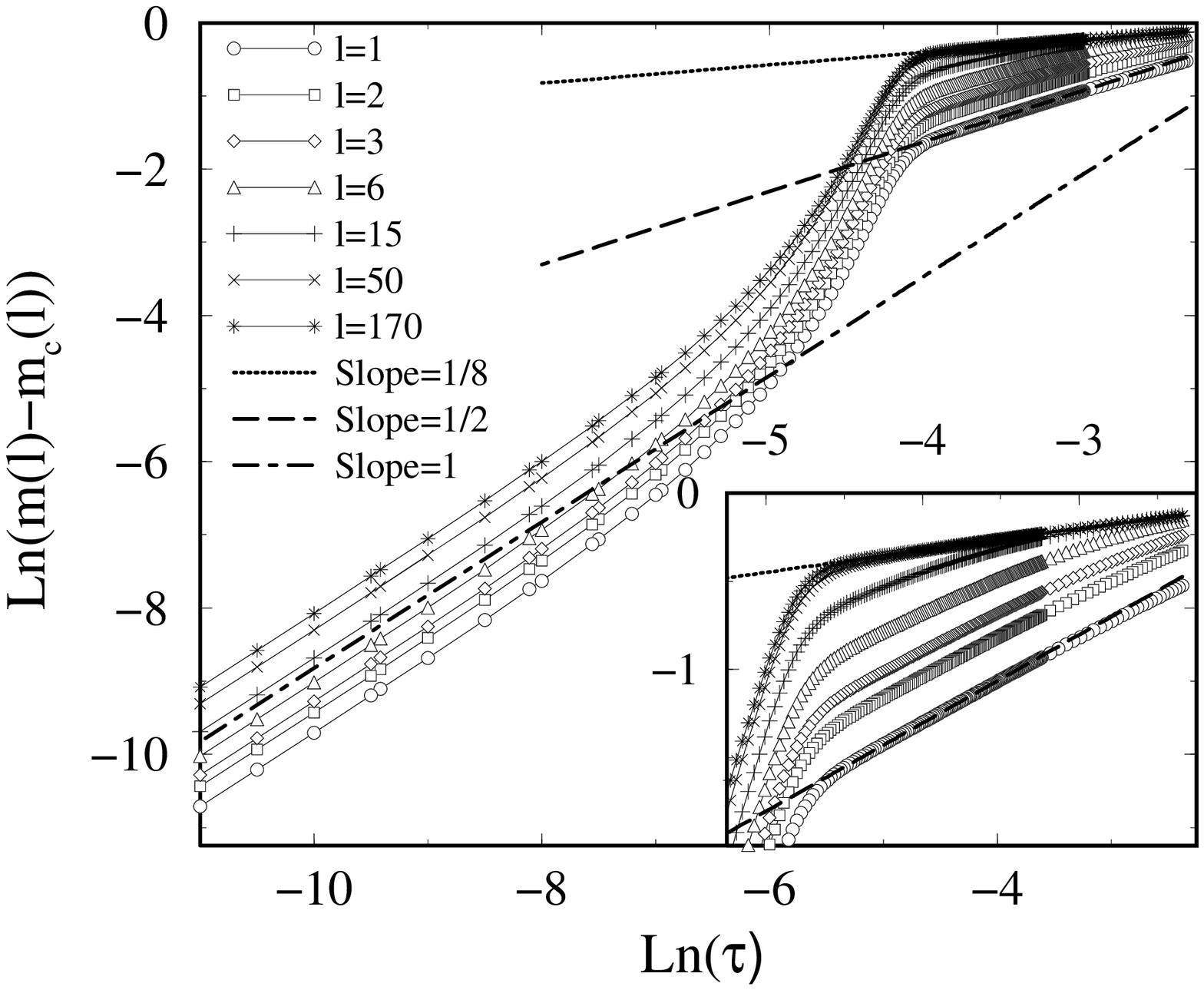}
\caption{The magnetization as a function of $\tau$ at $L=340$ for various 
strip layers. The inset presents the enlargement of the large-tau part.
}
\label{fig4}
\end{figure}

The crossover region is associated with
the formation of the maximum of the local exponent; it shrinks as the
surface field becomes stronger and disappears
altogether for $h_1=0.1$.
For the strongest considered $h_1$, i.e., for $h_1=0.5$ 
we find a linear behavior for $\tau < 0.005$;
again the amplitude of this leading decay depends on the surface field.
Our findings are consistent with the exact results (\ref{eq:mac6})
(\ref{eq:mac7}), $h_1\approx 0.1$ being the approximate  value of the
surface field for which one type of the limiting behavior crosses
over to another. 
We are not able to decide  whether the crossover to the linear
behavior observed for weak surface field  is  connected
with entering the supercritical region above the pseudocritical temperature
$T_{c,L}$.
Recall, that the shift of the critical point due to the finite wall
separation and the
symmetry-breaking boundary conditions is given by
(ignoring nonuniversal metric factors)
$\Delta T_c\equiv T_{c,L}-T_{c}\sim -L^{-1/\nu}X_c(h_1L^{\Delta_1/\nu}), 
\Delta h_c \sim -L^{-\Delta/\nu}Y_c(h_1L^{\Delta_1/\nu})$
with $\nu=1$,  $\Delta=1/15$
and $\Delta_1=1/2$ for $d=2$ Ising model \cite{fisher:1}. $X_c$ and $Y_c$
are universal scaling functions.  Mean-field analysis show that 
 the scaling function $X_c$ is of the order O(1) for small arguments and very
 weakly depends on the value of  $h_1$~\cite{fisher:1}. 
Thus in mean-field 
$\Delta T_c\approx -0.001$, but the scaling function
$X_c(\zeta)$, where $\zeta=h_1L^{\Delta_1/\nu}$,
is not known for   $d=2$ Ising magnet. 
In principle it can vary strongly with the argument.
Although the crossover depends sensitively on the value of the surface field, 
it is not connected with the wetting because  
$\tau _w\equiv (T_c-T_w(h_1))/T_c\simeq 8*10^{-7} $ for $h_1=0.001$,
$\simeq 10^{-5} $ for $h_1=0.005$, $\simeq 5*10^{-5}  $ for $h_1=0.01$,
$\simeq 5*10^{-4}  $ for $h_1=0.03$.  

Figs.~\ref{fig3}a and b show the  influence of the finite width
of the strip  on the critical behavior of  $m^I_1-m^I_{1c}$ for the weak
surface field, i.e., for $h_1=0.001$.
For  larger $L$ the crossover region
from the square-root to the linear behavior
is narrow and located  closer to $T_c$.
Notice, that for small systems 
 (see the curve for $L=50$ in Fig.~\ref{fig3}b)  the finite-size 
 effects are so strong that the ordinary transition behavior is attained
only very  far from $T_c$.

In the Fig.~\ref{fig4} we show that the magnetization
in the inside layers of the wide strip ($L=340$) subject to the weak
surface fields $h_1=0.001$ decays as  $\sim \tau^{\beta}$ with
$\beta=1/8$, i.e.,  as the  spontaneous magnetization,
and then crosses over to the linear dependence. 
 Similar behavior for the variation of
the local order parameter was presented by Brovchenko 
et al \cite{brovchenko:05}.
It is striking that
the crossover to the linear behavior takes place  in 
 approximately the same temperature range around  $~\sim 0.01$ for  all layers.

\section{Along the pseudo-coexistence}
\label{sec:2}

Now we consider the path  along the pseudo-coexistence of the
confined system $h=h_{co}(T;L,h_1)$.
When the Ising system is confined between parallel walls
subject to identical surface fields $ h_{1}$,  there is a shift of the
bulk first-order transition to a finite value of the bulk magnetic
field. In order to restore the coexistence the sign of the bulk
magnetic field $h$  has to be opposite  to the sign of $h_1$.
In $d=2$ for the system with finite $L$ there  is no unambiguous way to determine
the pseudocoexistence line. One can use several criteria, for example, 
maxima of the specific
heat, minima of the inverse correlation length or inflection
points of the solvation force \cite{maciolek:01}.  However, 
 above some characteristic 
temperature, which corresponds to the (pseudo)capillary critical point,
the curves based on different criteria separate because they
are governed by different critical exponent. Here we adapt a very natural
criterion  of the zero total magnetization,
i.e., for the fixed value of $h$  one  calculates the total magnetization
$\Gamma\equiv \sum _{l=1}^{L}m_l$, where $m_l$ is the magnetization in the
$l$-layer corresponding to a perpendicular distance from the first wall,
for different temperatures and identifies  $h=h_{co}(T,h_1)$ at the
temperature at which $\Gamma=0$. This method works
very good away from  the immediate neighborhood of the critical point, where
the difference between two phases vanishes and it is difficult to locate
the point corresponding to $\Gamma =0$. 
In the end we are not able to determine
the difference  $\Delta m_{1} \equiv m_1^I-m_1^{II}$ too
close to the bulk critical temperature (in the limit of $\tau \to 0$).

\begin{figure}[h]
\centering
\includegraphics[width=8.0cm]{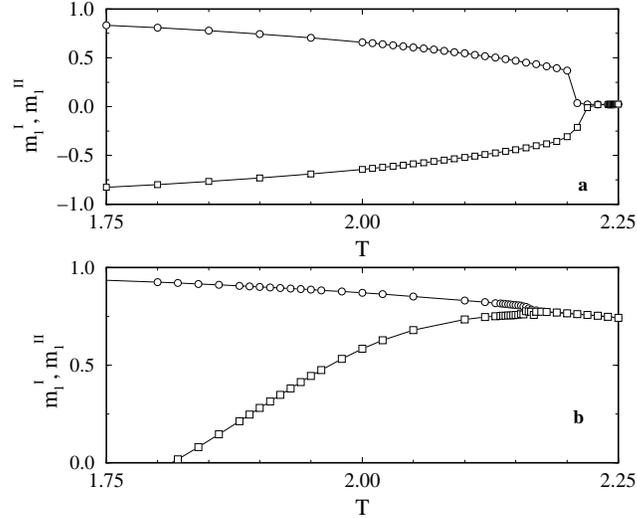}
\caption{The temperature dependence of the surface magnetizations of 
the two coexisting phases calculated along the (pseudo)coexistence 
lines: a) $h_1$=0.01 b) $h_1$= 0.5.
}
\label{fig5}
\end{figure}

We  have performed calculations for the strip of the width $L=340$ and
three different surface fields $h_1=0.01, 0.5, 0.8$.
The temperature dependence of the surface magnetizations of the two
coexisting phases $m^I_1$ and $m^{II}_1$ calculated along
the line $h=h_{co}(T,h_1)$ for $h_1=0.01$ and $0.5$
 are shown in the Fig.~\ref{fig5}. For the strongest field one
can  see the asymmetry due to the positive
surface field. The temperature at which these two curves coincide 
may be identified  with  $ T_{c,L}$.
Note, that the pseudocoexistence temperature is located approximately
at the same temperature at which the numerical errors become
relevant, i.e., for $h_1=0.01$ at $ \approx 2.22$, for $h_1=0.5$ at 
$ \approx 2.165$, ($T_c\approx 2.26919$ for the $d=2$ Ising model).
The appearence of the numerical errors is  connected with large fluctuations 
near
 $ T_{c,L}$

\begin{figure}[h]
\centering
\includegraphics[width=8.0cm]{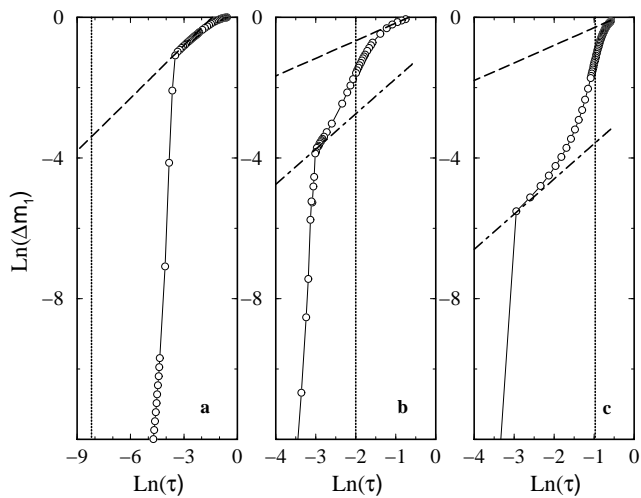}
\caption{The log-log plots of $\Delta m_1$ as a function of $\tau$ 
at L=340 for various surface fields: a) $h_1$=0.01 b) $h_1$=0.5 c) $h_1$= 0.8. 
The vertical dotted lines denote wetting temperatures, the dashed 
line presents the slope 1/2 and the dotted-dashed line
the slope 1.
}
\label{fig6}
\end{figure}

\begin{figure}[h]
\centering
\includegraphics[width=8.0cm]{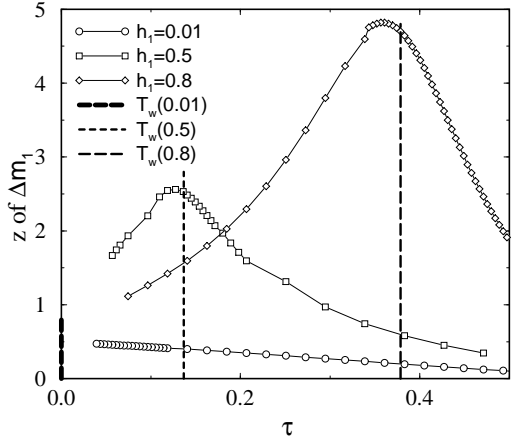}
\caption{The effective exponent of $\Delta m_1$ for various surface fields.
}
\label{fig7}
\end{figure}

The log-log plots of the difference  $\Delta m_{1} \equiv m_1^I-m_1^{II}$
versus $\tau$ for $h_1=0.01$, $h_1=0.5$ and $h_1=0.8$ are shown in 
Fig.~\ref{fig6}a,b,c, respectively, and the effective exponent
 is presented in Fig.~\ref{fig7}.
We also mark the wetting temperature $T_w(h_1)$.
The general behavior which can be read off from these plots
is that below $T_w(h_1)$ the difference  $\Delta m_{1} \equiv m_1^I-m_1^{II}$
decays approximately like  $\tau^{1/2}$, then there is a crossover regime
connected with the wetting transition, closer to $T_c$
the linear behavior dominates, and finaly there is a rapid
 decay due to the proximity of the pseudo-critical point.  For
$h_1=0.8$ the approximate square-root behavior takes place
 in a very narrow range of temperatures, because
$T_w$ lies quite far away from  $T_c$ ($(T_c-T_w)/T_w \approx 0.38$).
For $h_1=0.01$ the linear behavior is not reached since $T_w(0.01)\approx 2.26906$.

\begin{figure}[h]
\centering
\includegraphics[width=8.0cm]{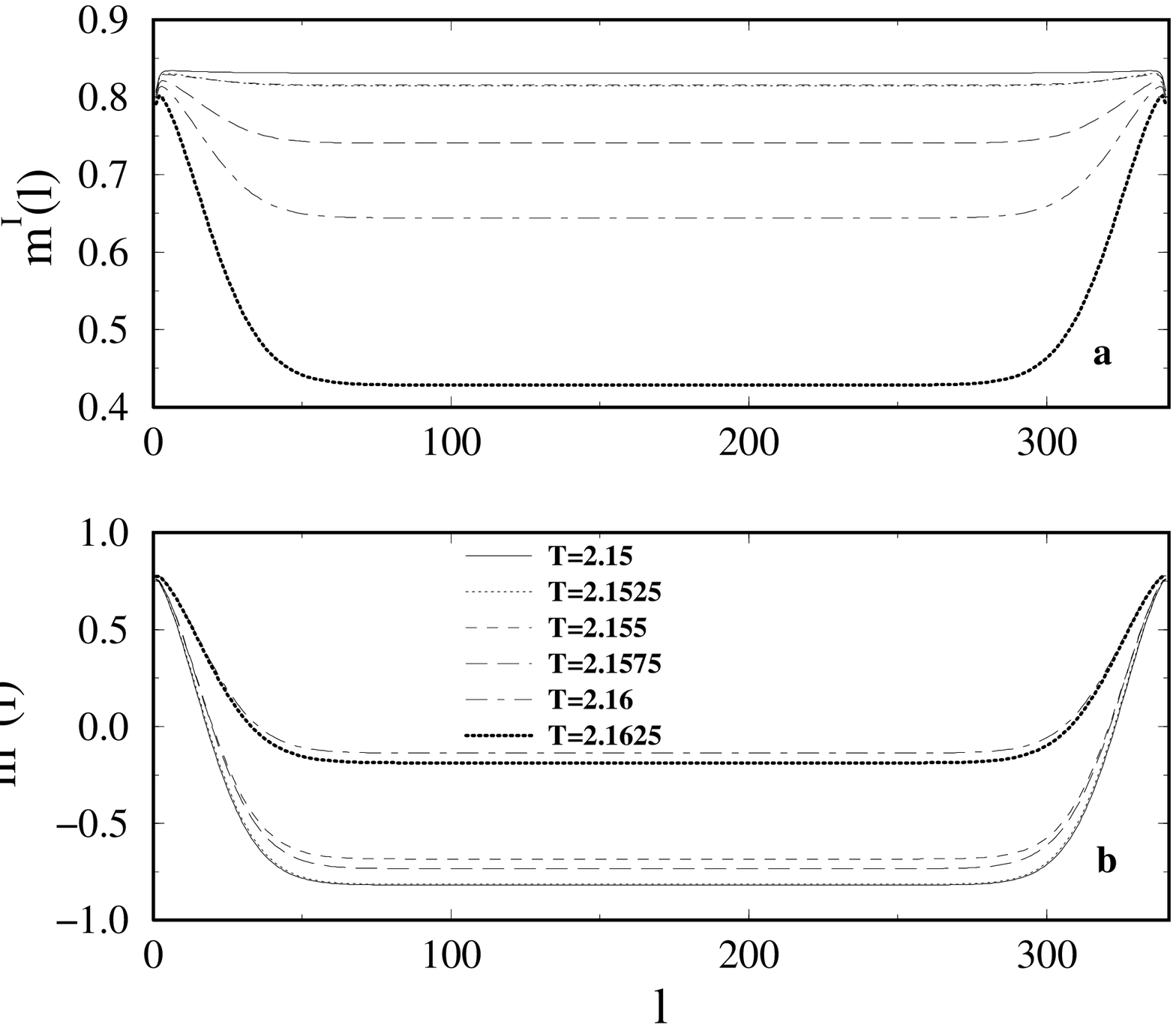}
\caption{The magnetization profiles near the (pseudo)critical point calculated
along the zero-magnetization line: a) $m^I$ b) $m^{II}$.
}
\label{fig8}
\end{figure}

\begin{figure}[h]
\centering
\includegraphics[width=8.0cm]{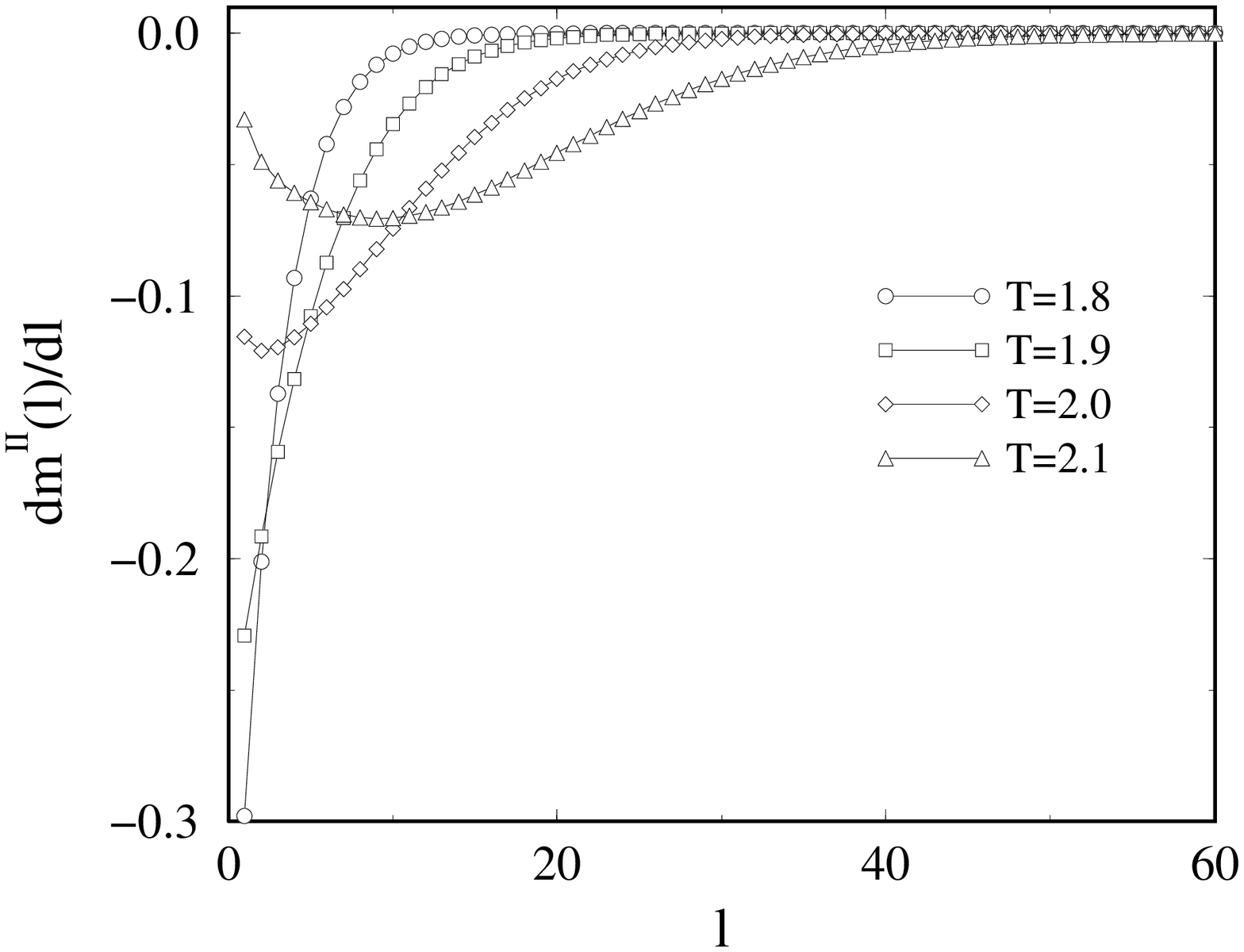}
\caption{The first derivative of the $m^{II}$ profiles with respect to the distance
from the surface at $h_1$=0.5. The strip width is L=340. The wetting temperature 
for this surface field is $T_w\sim1.96$.
}
\label{fig9}
\end{figure}

It is instructive to see the near surface behavior
of the magnetization profiles in both phases for different temperatures
(see Fig.~\ref{fig8}a,b). In the phase opposite to the one that is favored by
walls, i.e., in the negatively magnetized phase $m^{II}$,
the wetting transition manifests itself by a change in $dm^{II}(l)/dl$
from the monotonuous function of the distance from the surface $l$
to the one exhibiting a minimum (see Fig.~\ref{fig9}).

In conclusion, our DMRG results for the case of vanishing bulk
field  are in agreement with the exact results and  show two different
type of the asymptotic behavior of the surface magnetization. For weak
surface fields we see the square-root $\tau-$dependence characteristic of the
ordinary surface universality class which crosses over to the linear
behavior sufficiently close to $T_c$. This crossover is not 
 connected with the wetting. For $h_1 >  0.1$ we find the linear
behavior which dominates over the singular $\tau^{2-\alpha}$,
characteristic of the normal universality class.
Results along $h_{co}(T)$ suggest that $\Delta m_1 \sim
\tau^{1/2}$ below wetting transition but $\Delta m_1 \sim \tau$ above it.
However, in order to clear-out this issue the exact calculations of
$\Delta m_1$ in the {\it semiinfinite} system is needed. Because for very
weak surface field the wetting temperature lies very close to $T_c$
one may in such a case observe only the square-root behavior as in
the simulation of the Ref.~\cite{brovchenko:04,brovchenko:05}.

\end{document}